\documentclass[apj]{emulateapj}
\usepackage[ngerman,english]{babel}
\newif\ifAMStwofonts
\AMStwofontstrue

\def\cl{{C_l}}

\def\al{{a_{lm}}}

\def\gsim{~\rlap{$>$}{\lower 1.0ex\hbox{$\sim$}}}

\def\simpropto{\lower.2ex\hbox{$\; \buildrel \propto \over \sim \;$}}
\def\ltsim{\lower.5ex\hbox{$\; \buildrel < \over \sim \;$}}
\def\gtsim{\lower.5ex\hbox{$\; \buildrel > \over \sim \;$}}
\def\ltsim{\lower.5ex\hbox{$\; \buildrel < \over \sim \;$}}
\def\gtsim{\lower.5ex\hbox{$\; \buildrel > \over \sim \;$}}

\def\sun{\Theta_{_{\rm un}}}

\def\dd{\,{\rm d}}

\def\dd{{\rm d}}

\def\pmb#1{\setbox0=\hbox{#1}%
\kern-.025em\copy0\kern-\wd0
\kern.05em\copy0\kern-\wd0
\kern-.025em\raise.0433em\box0}

\def\vr{\pmb{r}}

\def\hvn{\hat {\vr}}

\def\vk{\pmb{k}}

\def\simlt{\lower.5ex\hbox{$\; \buildrel < \over \sim \;$}}
\def\simgt{\lower.5ex\hbox{$\; \buildrel > \over \sim \;$}}

\newcommand{\beq}{\begin{equation}}

\newcommand{\eeq}{\end{equation}}
\def\beqa{\begin{eqnarray}}
\def\eeqa{\end{eqnarray}}
\def\fixit#1{}

\def\dd{{\rm d}}

\def\apjs{ApJ. S}
\def\cN{{\cal N}}

\def\apjl{ApJL}

\def\aap{A\&A}

\def\apjl{ApJL}

\usepackage{bm}
\usepackage{color}
\usepackage{graphicx}
\usepackage{hyperref}
\usepackage{enumerate}
\begin{document}
\title{Large angular scale multipoles at redshift $\sim 0.8$}
\author{Prabhakar Tiwari }
\email{ptiwari@nao.cas.cn}
\affiliation{ National Astronomical Observatories, CAS, Beijing 100012, China}
\author{Pavan K. Aluri}
\email{pavanaluri.phy@itbhu.ac.in}
\affiliation{Department of Physics, Indian Institute of Technology (BHU), Varanasi 221005, India}
\date{\today}
\begin{abstract}
We prepare the full sky radio galaxy map ($|b|>10^{\circ}$) using the north NVSS and south SUMSS galaxy catalogs and study the large scale multipoles anomalies. These galaxies are roughly at redshift $z \sim 0.8$ and therefore tracing the matter distribution at very large scales. The quadruple and octopole from radio galaxy catalog are  consistent with $\Lambda$CDM for a reasonable value of galaxy bias and we do not find dipole--quadruple--octopole alignment as seen in CMB temperature maps. The quadrupole direction is roughly $46^{\circ}$ away from dipole, and octopole direction is approximately $33^{\circ}$ from dipole. The angle between quadrupole and octopole is around $70^\circ$ degree. We have large errors in multipole directions due to shot noise, even so with this data we are able to rule out dipole--quadruple and quadruple--octopole alignment. The magnitude of all  multipoles, except dipole, are roughly consistent with $\Lambda$CDM for reasonable galaxy bias. The dipole magnitude remains inconsistent with CMB as reported in previous studies. The results may impose stringent constraints on cosmological models with large scale anisotropy features.  
\end{abstract}
 \maketitle
\section{Introduction}
\label{sec:int}
In modern cosmology we assume our Universe at large scale to be statistically  homogeneous and isotropic \citep{Milne:1933CP,Milne:1935CP}. The Cosmic microwave background (CMB) is uniform to roughly 1 part in $10^5$ \citep{Penzias:1965,COBE_White:1994,WMAP:2013,Planck_iso:2016} and this strongly supports the isotropy assumption, furthermore there are other observations of isotropy e.g.  ultra-high energy cosmic ray (UHECR) events from the Telescope Array (TA) are  isotropic on the sky \citep{Abu-Zayyad:2012}, the Fermi Gamma-Ray Burst (GRB)  data is isotopic \citep{Ripa:2017,Ripa:2018} and the radio polarization angles from AGNs are also isotropic \citep{Tiwari:2018}. However, there remains several observations 
along with signals from CMB itself that suggest a violation of statistical isotropy.  In particular the CMB dipole, quadrupole and octopole modes are roughly aligned and are puzzling within the standard model of cosmology \citep{Schwarz:2004,Oliveira-Costa:2004,Schwarz:2016}. 
In the CMB map from nine years of Wilkinson Microwave Anisotropy Probe (WMAP) observations, quadrupole and octopole are aligned within  $3^\circ$ (degree) \citep{WMAP:2013}. Planck observations also confirm this result where quadrupole and octopole are found to be aligned at $8^\circ$ to $13^\circ$ in the foreground cleaned CMB maps produced by Planck team using various cleaning procedures. The probability of such an alignment to occur is $\sim 1$ to $2.6$ \% \citep{Planck_iso:2013} and thus indicates our assumption of isotropy at large scales.

The dipole--quadrupole--octopole alignment signal from CMB is unique as we have never been able to have alternate measure of this signal from some complimentary cosmological observation.
CMB anisotropies trace the density perturbations in the universe at redshift $\sim 1100$, when neutral Hydrogen was formed. The density perturbations grew giving rise to galaxies, galaxy clusters, and all the visible/non-visible cosmological structure around us.
The high density sites of dark matter, the halos, mediated the baryonic matter to form galaxies, and so the galaxy distribution in space is tracing the background dark matter with galaxy {\it bias} \citep{Kaiser:1984}. Therefore with the large-scale galaxy surveys, we can probe the background dark matter distribution and can test for any large-scale anomalies in it.
Any alignment thus observed in large scale multipoles of the galaxy distribution map will constitute an independent measure of a similar feature in background dark matter distribution that is so far seen (only) in CMB.
The galaxies in this work are sitting at redshift around $\sim 0.8$ \cite{Condon:1998,Wilman:2008} and so we will be probing the anomalies, if any, in the background dark matter distribution at this redshift. Nevertheless, it is worth looking at how the observed CMB anomalies (at redshift $\sim$1100 ) got transformed with structure formation and how these alignments look like in galaxy surveys, if the anomalies are truly cosmological.

The NRAO VLA Sky Survey (NVSS) radio galaxy clustering results, for higher order multipoles ($l>4$) show an 
excellent consistency with the standard $\Lambda$CDM power spectrum for a reasonable choice of {\it bias} parameter \citep{Blake:2002ac,Adi:2015nb}. However the dipole signal from NVSS galaxies is significantly higher, roughly three times larger than CMB predicted value\footnote{Our local motion with respect to CMB frame is observed as dipole signal in CMB temperature map, which is of the order of few milli Kelvin \citep{Conklin:1971,Henry:1971,Corey:1976,Smoot:1977,Kogut:1993,Hinshaw:2009} and 
corresponds to a speed of 369$\pm$ 0.9 km s$^{-1}$ in the direction $l=263.99^o\pm 0.14^o$, $b=48.26\pm 0.03^o$ 
in galactic coordinates \citep{Kogut:1993,Hinshaw:2009}. Therefore we also expect a dipole signal in galaxy distribution due to Doppler and aberration effects \citep{Ellis:1984} caused by our local motion.} \citep{Singal:2011,Gibelyou:2012,Rubart:2013,Tiwari:2014ni,Tiwari:2015np,Tiwari:2016adi}.

In this work we aim to study the dipole, quadrupole, and octopole modes, and their alignments from radio galaxy catalogs. 
NVSS covers the sky north of declination $-40^\circ$ (J2000), which is almost 80\% of the celestial sphere, however 
the remaining 20 \% southern sky remains as an obstacle to achieve a confident measure of large-scale multipoles 
i.e. quadrupole and octopole. Thankfully, we have Sydney University Molonglo Sky Survey (SUMSS) in southern sky and 
by merging NVSS and SUMSS we achieve full sky radio galaxy map for latitudes $|b|>10^{\circ}$ \citep{Colin:2017}. 
We carefully prepare the NVSUMSS near full sky galaxy catalog and estimate all the multipoles in the number density map. Our NVSUMSS merging method and final catalog differs from \cite{Colin:2017}. 
Assuming bias value determined in \cite{Adi:2015nb} we compare the estimated multipoles with $\Lambda$CDM. Next, we employ the Power tensor method \citep{Ralston:2004,Samal:2008} to determine the direction of large scale multipoles and study their alignments. 

Outline of the remaining paper is as following. We discuss the data and full sky catalog preparation in Section \ref{sec:data}. In Section \ref{sec:theory}, we review the angular power spectrum, $C_l$, formulation and discuss its estimation from the data with partial sky coverage 
($|b|>10^{\circ}$). The clustering results recovered with NVSUMSS are presented in Section \ref{sec:cal}. A comparison of angular clustering results obtained using two different methods, and their matching and calibration 
with previous studies is also presented in this section. The dipole--quadrupole--octopole alignment analysis and results are presented in Section \ref{sec:l23}. We conclude with a discussion of our results in Section \ref{sec:con}.

\section{Full sky radio galaxy catalog}%
\label{sec:data}

\subsection{NVSS}
\label{ssc:nvss}
The NVSS\footnote{\href{https://www.cv.nrao.edu/nvss/}{https://www.cv.nrao.edu/nvss/}} catalog covers the sky north of declination $-40^\circ$ (J2000) in Equatorial coordinates. This is almost 80$\%$ (69\% if we mask $|b|<10^{\circ }$) of the celestial sphere. The full catalog contains $\sim$1.7 billion sources with integrated flux density $S>2.5\rm ~mJy$ at 1.4 GHz and it is approximately complete above $3.5\rm ~mJy$ \citep{Condon:1998}. The full width at half maximum 
resolution of the survey i.e.  FWHM  is $45''$ (arcsec) and observations are at nearly uniform sensitivity. The catalog is known to have some 
systematics namely the Galactic contamination, 22 bright extended source locations  and significant systematic gradients in surface density for  sources fainter than $10 ~\rm mJy$ due to array  D and  DnC configuration \citep{Blake:2002}. After masking 22 bright extended source sites and Galactic sources with  latitudes $|b|<5^{\circ }$ the galaxy spatial distribution is reasonably smooth with flux density cut $S = 10$ mJy and above \citep{Blake:2002ac,Adi:2015nb}.

\subsection{SUMSS}
\label{ssc:sumss}
The SUMSS\footnote{\href{http://www.physics.usyd.edu.au/sifa/Main/SUMSS}{http://www.physics.usyd.edu.au/sifa/Main/SUMSS}} catalog covers the sky south of declination $-30^\circ$ (J2000). The survey is carried out with 
Molonglo Observatory Synthesis Telescope (MOST) operating at 843 MHz \citep{Mauch:2003}. The catalog 
is limited to Galactic latitudes $|b|>10^{\circ }$.
The catalog is complete above $ 8 \rm ~mJy$ at declination $\leq -50$ and for declination between $-50$ 
to $-30$ it is complete above 18 mJy at 843 MHz. The survey is uniform over the observation region and with similar FWHM resolution and sensitivity to NVSS. 

\subsection{NVSUMSS}
\label{ssc:nvsumss}
The NVSS and SUMSS operate at different frequencies and thus for a given source the radio flux measurements are different. Nevertheless the radio fluxes at these two frequencies can be linked using the relation,
\beq
S \propto \nu^{-\alpha},
\eeq
where $\alpha$ is the spectral index. Therefore for a given source,
\beq
\label{eq:f_scale}
S_{1.4 \rm~GHz} = S_{843{\rm~ MHz}} (843/1400)^{\alpha}.
\eeq

\begin{figure}
 \includegraphics[width=0.5\textwidth]{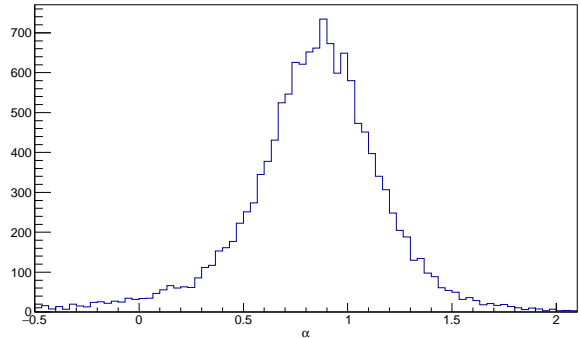}
 \caption{The spectral index with sources common to NVSS and SUMSS. There are 13942 common sources and spectral index, $\alpha=0.83\pm0.35$.}
\label{fig:alp}
\end{figure}

The two surveys, NVSS and SUMSS, have an overlap region between sky south of declination $-40^\circ$ to $-30^\circ$. We employ this common survey region to obtain the spectral index $\alpha$. Considering NVSS and SUMSS position uncertainties we cross match these catalogs.  The cross matching of catalogues is done as described below.

We have in total 35579 sources in SUMSS above 18 mJy in the overlap region of the two surveys. We do find at least one source position match from NVSS for most of the SUMSS sources (35502 sources out of 35579) if  we consider $45''$ (arcsec) error in RA \& Dec for all SUMSS's source positions, i.e., looking for a NVSS source within a circle of $45''$ radius around each SUMSS's source. However if we use the source position uncertainties in RA \& Dec as provided in the catalog, we only match a total of 13942 positions. The spectral index distribution obtained using the common sources (13942) between these catalogs following equation \ref{eq:f_scale} is shown in Figure \ref{fig:alp}. We find $\alpha=0.83\pm0.35$.

Alternately we can also obtain an estimate of the spectral index as follows.
In the common observing region of the two surveys, SUMSS is complete above 18 mJy at 843 MHz and NVSS is complete above 3.5 mJy at 1.4 GHz (note that 18 mJy at 843 MHz corresponds to 12 mJy at  1.4 GHz following equation \ref{eq:f_scale}). So for a given flux cut which is above the flux completeness limit of both surveys, we expect to see same radio galaxies and thus the number density must match. Indeed, we recover the same number density ($\pm$0.3 \%) from these surveys with $\alpha \approx 0.81$. This value of $\alpha$ is slightly less than the mean spectral index from Figure \ref{fig:alp}. This tiny deviation may occur as these surveys are at low resolution and dominated by unresolved sources.

\begin{figure}
\includegraphics[width=0.5\textwidth]{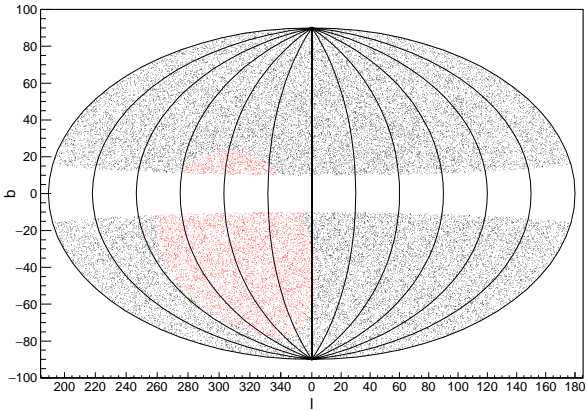}%
\caption{The NVSUMSS above $S>15\rm mJy$ at 1.4 GHz, the SUMSS flux density scaled with $\alpha=0.81$. The NVSS (black) covers the sky north of declination $-40^\circ$ (J2000) and SUMSS (red) fills the remaining sky. The Aitoff projection is in the galactic coordinate. }
 \label{fig:NVSUMSS}
\end{figure}

In our present work we use $\alpha=0.81$ to scale the SUMSS observed source fluxes from 843 MHz to 1.4 GHz, following equation \ref{eq:f_scale}, and combine them with NVSS sources. We remove Galactic plane $|b|<10^{\circ }$ and also 22 bright extend sources and produce our otherwise full sky radio galaxy map, the ``NVSUMSS''. It spans about 82.74\% of the sky and contains 410308 sources above 15 mJy at 1.4 GHz. The NVSUMSS is complete above 12 mJy at 1.4 GHz (since 12 mJy at 1.4 GHz corresponds to 18 mJy at 843 MHz - the SUMSS completeness limit). NVSUMSS catalog in Aitoff projection in the galactic coordinates is shown in Figure \ref{fig:NVSUMSS}.

\section{The galaxy clustering angular power spectrum}
\label{sec:theory}
\subsection{Theoretical $\cl$}
Here we briefly review the relationship between the galaxies' spatial distribution and background dark matter following $\Lambda$CDM scenario. 
Let $\cN(\hvn)$ be  the projected number density per steredian in  the direction $\hvn$. We can write this as,
\beq
\cN (\hvn)=\bar \cN(1+\Delta (\hvn)),
\eeq
where $\bar \cN$  be the mean number density, and  $\Delta (\hvn)$ represents 
the projected number density contrast. $\Delta (\hvn)$ is theoretically connected to the background dark matter density contrast, $\delta_m(\vr,z(r))$. Here $\vr$ stands for comoving distance $r$ in direction $\hat{r}$ and $z(r)$ is the redshift corresponding to comoving distance $r$. Assuming linear galaxy biasing $b(z)$ we can write the galaxy density contrast,
\beq
\delta_g(\vr,z(r)) =\delta_m(\vr,z=0) D(z) b(z),
\eeq
where $D(z)$ is the linear growth factor and $z=z(r)$. Following these we get,
\begin{eqnarray}
\label{eq:delta_th}
\Delta (\hvn) &=& \int _{0}^{\infty} \delta_g(\vr, z(r)) p(r) dr  \nonumber\\
              &=&  \int _{0}^{\infty} \delta_m(\vr,z=0)  D(z) b(z) p(r) dr , 
\end{eqnarray}
where  $p(r) \dd r$ is the probability of observing galaxy between comoving distance $r$ and
$(r+ dr)$.  Next we can expand  $\Delta (\hvn)$ in spherical harmonics as,
\beq
\label{eq:alm}
\Delta (\hvn) = \sum_{lm}a_{lm}Y_{lm}(\hvn).
\eeq
We invert the above equation to recover $\al$ as,
\begin{eqnarray}
\label{eq:alm_gal}
\al&=&\int d \Omega \Delta(\hvn)  Y_{lm}^*(\hvn)\\
\nonumber &=& \int d\Omega Y_{lm}^*(\hvn) \int_{0}^{\infty} \delta_m(\vr,z=0)  D(z) b(z) p(r) dr\;.
\end{eqnarray}
The dark matter density field $\delta_m(\vr,z=0)$ can be written as a Fourier transform of $k$-space density field $\delta_{\vk}$, as
\begin{equation}
\delta_m(\vr,z=0) =\frac{1}{(2\pi)^3}\int d^3 k \delta_{\vk}{\rm e}^{i \vk \cdot{  \vr}}\;.
\end{equation}
Here we can substitute, 
\beq
\label{eq:ekr}
{\rm e}^{i \vk \cdot{  \vr}}=4\pi \sum_{l,m} {i}^lj_l(kr) Y^*_{lm}(\hat {\vr})Y_{lm}(\hat {\vk}),  
\eeq
where $j_l$ is the spherical Bessel function of first kind for integer $l$. Subsequently we write
\begin{equation}
\label{eq:alm_th2}
\al=\frac{{i}^l}{2\pi^2}\int dr D(z) b(z) p(r) \int d^3 k \delta_{\vk}j_l(kr) Y^*_{lm}(\hat {\vk}) \; . 
\end{equation}

Now we can obtain an expression for the corresponding angular power spectrum, $C_l$, as
\begin{eqnarray}
\label{eq:clth}
C_l&=&<|\al |^2> \nonumber\\
\nonumber &=& \frac{2}{\pi }\int dk k^2 P(k) \left\vert \int_{0}^{\infty} D(z) b(z) p(r) d r  j_l(kr)\right\vert^2 \\
          &=&   \frac{2}{\pi }\int dk k^2 P(k) W^2(k) \;,
\end{eqnarray}
where $P(k)$ is $\Lambda$CDM power spectrum, and  $W(k)=\int_{0}^{\infty} D(z) b(z) p(r) d r  j_l(kr)$ is the window function in $k$-space. We have used $<\delta_{\vk}\delta_{\vk'}>=(2\pi)^3 \delta^{\rm D}(\vk-\vk')P(k)$ where $\delta^{\rm D}$ is Dirac's $\delta-$function.

\subsection{Measured $C_l$ from galaxy surveys}
\label{ssc:clobs}
The galaxy surveys in reality never cover the full sky and we have some regions with no data or bad data in sky.
Therefore in general the measured $C_l$ are always from partial sky. Furthermore we are limited by galaxy number density and thus we have shot noise in $C_l$ measurements. An estimate of $C_l$ corresponding to the theoretical $C_l$ given in Equation (\ref{eq:clth}) is,
\begin{equation}
\label{eq:cobs}
C^{\rm obs}_l=\frac{\langle  |a^{\prime}_{lm}|^2\rangle}{J_{lm}} -\frac{1}{\bar \cN}
\end{equation}
where  $a^{\prime}_{lm} =\int_{\rm survey} d \Omega  \Delta(\hvn) Y_{lm}^*(\hvn)$  and
$J_{lm}=\int_{_{\rm survey}}|Y_{lm}|^2 \dd \Omega$, the $J_{lm}$ is the approximate
correction for the partial survey region following \cite{Peebles:1980}. The term $\frac{1}{\bar \cN}$ is deducted to remove the contribution from the Poissonian shot noise. The 1$\sigma$ error in this estimate due to cosmic variance, sky coverage and shot-noise is,
\begin{equation}
\label{eq:dcl}
\Delta C_l = \sqrt{\frac{2}{(2l+1) f_{\rm sky}}} \left(C^{\rm obs}_l + \frac{1}{\bar \cN}\right)
\end{equation}
where $f_{\rm sky}$ is the fraction of sky observed in the survey.

\section{Clustering power spectrum from NVSUMSS}
\label{sec:cal}
\begin{figure}
        \includegraphics[width=0.48\textwidth]{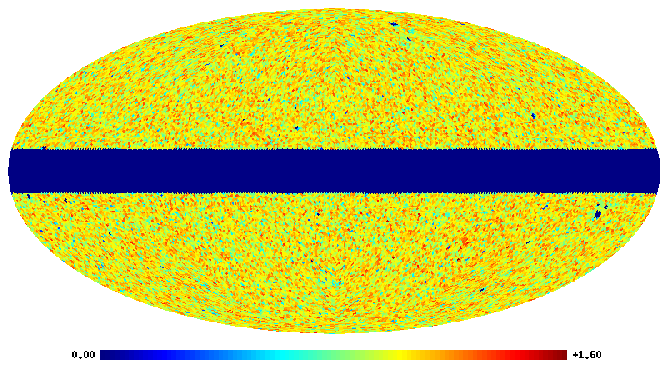}\\
        \includegraphics[width=0.48\textwidth]{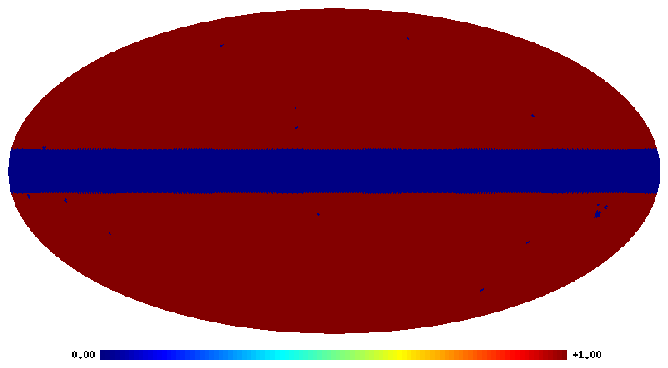}
        \caption{
        \emph{Top:} The number density log$_{10}$(1+N) map obtained using NVSUMSS with S$>15$ mJy at 1.4 GHz.  Here N denotes the number of sources in Healpix pixels. We have used N$_{\rm side} =64$. \emph{Bottom:} The corresponding source mask that denotes the extent of the sky covered by the composite catalog.}
        \label{fig:num-dens-map-mask}
\end{figure}

We use HEALPix\footnote{\href{https://healpix.jpl.nasa.gov/}{https://healpix.jpl.nasa.gov/}} 
\citep{Gorski:2005} pixelization scheme to produce equal area pixels on spherical surface. 
We next populate the map with our NVSUMSS  catalog and this gives the number density map $\cN (\hvn)$ i.e. the number of sources in a pixel in direction $\hvn$. We use an $N_{side}=64$ HEALPix grid to generate our number density map. The map thus obtained and the source mask are shown in Figure~\ref{fig:num-dens-map-mask}.

We obtain $C^{\rm obs}_l$ and its error bars following Equation \ref{eq:cobs} and \ref{eq:dcl}, respectively. 
The NVSS and SUMSS have broad angular resolution and around 90\% sources are unresolved in these surveys. The radio-loud sources often have extended radio emission and the resolved 10\% population may have multiple entries in the catalog. \cite{Blake:2004} noted that this has a small but measurable effect on angular power spectrum as a fixed offset to $C_l$ given by $\Delta C_l \approx 2 e/\bar{\cN}$, where $e=0.070\pm 0.005$. Note the factor $1/\bar{\cN}$ which is the shot noise contribution. Thus we also deduct this fixed offset from all $C_l$ in estimating $C_l^{\rm obs}$.

\begin{figure}
	\includegraphics[width=0.5\textwidth]{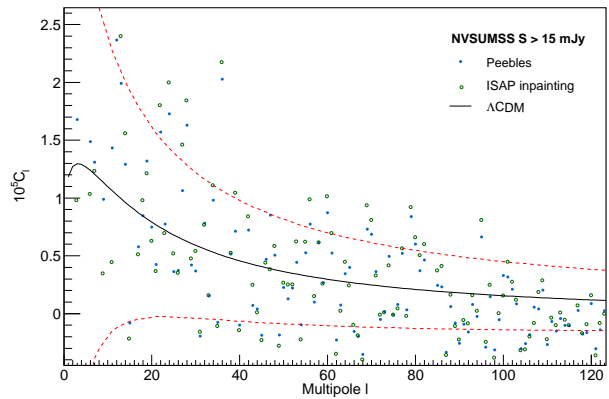}%
        \caption{The angular power spectrum estimated from the NVSUMSS catalog with flux density cut S$>$ 15 mJy at 
	1.4 GHz. The blue filled circles are $C^{\rm obs}_l$ following Equation \ref{eq:cobs}. The solid line is 
	$\Lambda$CDM values. The dashed lines are one sigma limits due to shot noise and cosmic variance scatter
	and $f_{\rm sky}$ (Equation \ref{eq:dcl}). The open circles are $C^{\rm obs}_l$ recovered using iSAP 
	inpainting scheme \citep{Starck:2013,Fourt:2013}.}
        \label{fig:clobs}
\end{figure}

\begin{figure}
 \includegraphics[width=0.5\textwidth]{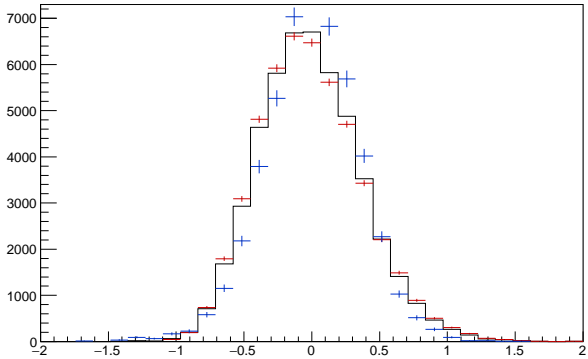}
 \caption{ The NVSUMSS number density contrast map histogram as seen in Healpix pixels. The inpainted full sky distribution  is shown in solid black and the input partial sky distribution is shown as blue. The recovered inpaited pixels distribution is shown in red. The input sky distribution and inpainted pixels histograms are scaled by a factor $49152/40667$ and $49152/8485$, respectively,  for comparison between the histograms.}
\label{fig:nvsumss-dist-inp}
\end{figure}

The $C^{\rm obs}_l$ from NVSUMSS are shown in Figure \ref{fig:clobs}. The NVSUMSS $C_l$ agree with $C_l$ obtained using only NVSS sources by \cite{Adi:2015nb}. The solid curve in  Figure~\ref{fig:clobs} denotes the $\Lambda$CDM angular power spectrum following the bias and N(z) schemes given in \cite{Adi:2015nb}. We conclude that the clustering results from NVSUMSS are similar to NVSS and agrees well with $\Lambda$CDM predictions. This confirms that our NVSUMSS catalog is free from systematics and unusual clustering.
Since the main aim of the paper is to study the large angle multipoles i.e., low-$l$ modes
we use the iSAP inpainting package \citep{Starck:2013,Fourt:2013} to construct full sky $\al$ from the partial sky map surface number density contrast map shown in Figure~\ref{fig:num-dens-map-mask}. 
The default setting of the iSAP inpainting package were used to inpaint the missing portions of the sky. We have shown the inpainted full sky pixel distribution along with input partial map in Figure~\ref{fig:nvsumss-dist-inp}. The power spectrum from the inpainted number density map is also shown in Figure~\ref{fig:clobs} along with the one estimated using the classical method of \cite{Peebles:1980}. The power spectra recovered from both the methods match reasonably well with each other within the error bars. We note that, at low-$l$, the error bars in $C_ls$ due to cosmic variance are large.

\section{Dipole--quadrupole--octopole alignment analysis}
\label{sec:l23}

\subsection{Power tensor}
\label{ssc:powertensor}
In order to test for alignments among various multipoles we use the Power tensor 
method introduced in \cite{Ralston:2004}. The method involves associating a preferred 
axis or axis of anisotropy with each multipole. Then their orientations can be compared 
by taking a simple inner product of the preferred axis associated with the multipoles 
in question. For a range of multipoles, one can also compare their alignments 
using what is called `Alignment tensor' given by \cite{Samal:2008}.

Power tensor is defined as a quadratic estimator in the spherical harmonic 
coefficients, $\al$, of a multipole `$l$' as
\beq
A_{ij}(l) = \sum_{m,m'm''}a_{lm}J^i_{mm'}J^j_{m'm''}a_{lm''}
\eeq
where $J_i$ (i=1,2,3) are the angular momentum matrices in spin$-l$ representation.
The normalization factor is chosen such that the trace of this $3\times3$ Power tensor 
matrix corresponding to a multipole, $l$, is equal to the total power, $C_l$, of that multipole.

The Power tensor $A_{ij}(l)$ maps a multipole, or anologously $\al$s, to an ellipsoid.
Let $\Lambda_\alpha$ and $\bf{e}_\alpha$ denote the three eigenvalues and eigenvectors 
of the Power tensor respectively. These eigenvectors form the three perpendicular 
axes of the ellipsoid and the corresponding (normalized) eigenvalues denote 
length of each axis. We associate that eigenvector, which has the largest eigenvalue 
among the three $\Lambda_\alpha$'s, as the preferred axis or axis of anisotropy 
of a multipole $l$. Stated differently, the axis along which the ellipsoid is 
most elongated is taken as the axis of anisotropy of that multipole. We call this 
axis as Principal eigenvector (PEV) of that multipole. 
In the case of statistical isotropy all the eigenvalues of $A_{ij}(l)$ will 
be equal to $C_l/3$ and the PEVs will be oriented randomly.

For further details about the Power tensor method the reader may refer to \cite{Ralston:2004,Samal:2008}.

\subsection{Alignment analysis with Power tensor}
\label{ssc:algn-results-pt}
\begin{figure}
        \includegraphics[width=0.5\textwidth]{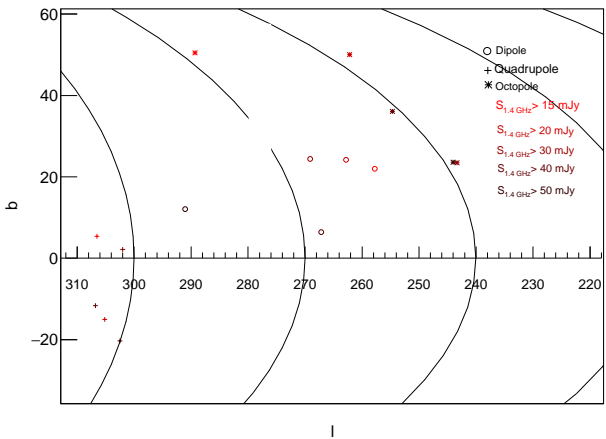}%
        \caption{Dipole, quadrupole and octopole directions from NVSUMSS in galactic coordinate for flux density cut S$>$ 15, 20, 30, 40, 50 mJy.}
        \label{fig:l123_dir}
\end{figure}

The amplitude of dipole as observed with NVSS and NVSS+SUMSS remains high and disagrees with CMB kinematic dipole \citep{Singal:2011,Gibelyou:2012,Rubart:2013,Tiwari:2014ni,Tiwari:2015np,Tiwari:2016adi,Colin:2017}. The magnitude of quadrupole is roughly at one sigma away from $\Lambda$CDM prediction and octopole magnitude is almost same as $\Lambda$CDM prediction. The observed power, $C^{\rm obs}_l$, for $l=1,2,3$ i.e. dipole, quadrupole and octopole modes are given in Table \ref{tb:l123}. The preferred direction i.e. PEV direction inferred for these multipoles using Power tensor are also listed in the same Table. We recall that these multipole directions are derived using spherical harmonic coefficients of the full-sky projected number density map obtained using iSAP 
inpainting procedure. The multipole power and preferred direction results for $l=1,2,3$ with different flux density cuts are given in Table \ref{tb:scut}. 
We note that the dipole direction and magnitude obtained with NVSUMSS is 
matching with previous studies  \citep{Tiwari:2016adi,Colin:2017}. 
The dipole, quadrupole and octopole  directions varies with flux density cuts but roughly remains stable. These multipole directions for various flux 
density cuts are also shown in Figure \ref{fig:l123_dir}.

The angle between dipole, quadrupole and octopole for various flux density cuts are given in Table \ref{tb:ang_diff}. The average  angle between dipole and quadrupole is $\approx 46^\circ$ and with octopole it is $\approx 33^\circ$. The quadrupole and octopole are $\approx 70^\circ$ away from each other on average. If we assume that the quadrupole and octopole directions are random and have no alignment with dipole then the probability of observed alignments are 0.30 and 0.16 for quadrupole--dipole and octopole--dipole, respectively. The probability of having quadrupole and octopole within $70^\circ$ is 0.66. This is statistically consistent with random direction assumption for multipoles.
Note that the multipole directions are axes and their orientation is random over the sky. The angle between two multipoles can be at maximum $90^\circ$ and probability of having two multipoles aligned within angle $\alpha$ is $\int^{\alpha}_{0} sin(\theta) d\theta$.   
\begin{table}
	\caption{ Dipole, quadrupole, octopole magnitude and direction from NVSUMSS with 
	flux density cut S$>$ 15 mJy at 1.4 GHz. The shot noise, $\frac{1}{\bar \cN}$, for this flux cut is 
	$2.5 \times 10^{-5}$.}
\centering
  \begin{tabular}{| c |c | c | c | c | c |}
    \hline
        $l$ & \multicolumn{3}{c} {$C^{\rm obs}_l$ ($\times 10^{4}$) } & \multicolumn{2}{c}{Direction (iSAP)}\\%
            & $\Lambda$CDM &  Peebles  &  iSAP  &  (l,b)& (RA, Dec) \\
   \hline 
	  1 & 0.127 &  2.141 & 2.234  & 253, 19 & 141, -23 \\
	  2 & 0.130 &  0.776 & 0.559  & 306, ~5 & 199, -58   \\
	  3 & 0.129 &  0.168 & 0.098  & 266, 46 & 168, -10\\
  \hline
  \end{tabular}
	\label{tb:l123}
\end{table}

\begin{table}
        \caption{ Dipole, quadrupole, octopole direction from NVSUMSS with 
        flux density cut S$>$ 20, 30, 40, 50 mJy. The shot noise, $\frac{1}{\bar \cN}$, for corresponding flux cuts
	is also given.}
\centering
  \begin{tabular}{| c |c | c | c | c |}
    \hline
	 $S$ &        \multicolumn{3}{c} {(l,b)}             & shot noise         \\%
    ($>$mJy) &  $l=1$         &  $l=2$       & $l=3$         &  ($\times 10^{5}$) \\
   \hline
	  20 & 258,  21 & 304, -14  & 235, 42 &  3.2\\
	  30 & 264,  22 & 302, ~~2  & 238, 19 &  4.7\\
	  40 & 267,  ~6 & 306, -11  & 242, 30 &  6.2\\
	  50 & 290,  11 & 300, -19  & 238, 20 &  7.8\\
  \hline
  \end{tabular}
        \label{tb:scut}
\end{table}
\begin{table}
        \caption{ Angular distance between dipole, quadrupole, octopole directions from NVSUMSS with 
        flux density cut S$>$15,20, 30, 40, 50 mJy. Last column is the average (over flux density cuts) angle between multipoles.}
\centering
  \begin{tabular}{| c |c | c | c |c|c|c|c|}
    \hline
	 Multipoles &\multicolumn{5}{c} {angle between (in degrees)} & Mean  \\%
       &S$>$15 & S$>$20 & S$>$30 &S$>$40 &S$>$50& $ $\\
   \hline
	  $l12$ & 54 & 58 & 42 & 43 & 32 & \color{red}46\\
	  $l13$ & 29 & 28 & 25 & 34 & 50 & \color{red}33\\
	  $l23$ & 54 & 85 & 65 & 76 & 72 & \color{red}70 \\
	  
  \hline
  \end{tabular}
        \label{tb:ang_diff}
\end{table}
\subsection{Multipole direction error estimate}
\label{ssc:algn-err}
In Figure \ref{fig:l123_dir} we have shown the multipole directions with different flux density
cuts. The scatter in multipole directions reflect the effect of shot noise, partial to full sky construction, etc.
To determine the effect of masking and shot noise, we resort to mocks and emulate the NVSUMSS multipole recovery  as following. We consider the NVSUMSS density contrast map, $m0$, and the multipole directions from this density contrast map as our model and input directions respectively. The NVSUMSS galaxy density map, $m$, with different flux density cuts, $S>$ 15 to 50 mJy, i.e. with different number density is simply $m(i)=(m0(i)+1)\times \bar{p}$, where $i$ stands for pixel and $\bar{p}$ is the mean number of galaxies in pixel i.e. pixel area $\times$ number density ($\bar{\cN}$). We call Poisson distribution at every pixel and prepare a new map, $m_{\rm poi}$. The pixels in $m_{\rm poi}$ map are given by, $m_{\rm poi} (i)=$Poisson ($m(i)$). Note that the map $m_{\rm poi}$ contains both the shot noise (i.e. 1/$\bar{\cN}$ ) and the model map $m$.
To emulate the effect of partial to full sky recovery we
employ the same NVSUMSS mask and use iSAP inpainting package to recover full sky map from partial $m_{\rm poi}$. We call this new map as $m'_{\rm poi}$ map. The $m'_{\rm poi}$ includes shot noise and partial to full sky recovery uncertainties. Next we construct density contrast map from $m'_{\rm poi}$ and apply Power tensor method to obtain multipole directions. The multipole magnitude and directions thus recovered, contain the model input map, $m0$, shot noise and also include uncertainties due to partial to full sky construction.  
We average over a large number of Poissonian maps following the same NVSUMSS mask and determine the uncertainty in multipole directions.  The results are listed in Table \ref{tb:dir_err}. We note that the direction recovery for the NVSUMSS octopole for flux cut $S>$ 15 and 20 mJy show a very large error in longitude. 

\begin{table}
        \caption{ Expected uncertainty in dipole, quadrupole, octopole direction due to shot-noise and
        partial to full sky recovery with flux density cut S$>$ 20, 30, 40, 50 mJy. Note the
        shot noise for these flux cuts in Table \ref{tb:scut}.}
\centering
  \begin{tabular}{| c |c | c | c |}
    \hline
         $S$ &        \multicolumn{3}{c} {($\Delta$l, $\Delta$b)}            \\%
    ($>$mJy) &  $l=1$         &  $l=2$       & $l=3$        \\
   \hline
          15 &16, 12 & 18, 17 &  65, 23  \\
          20 &14, 11 & 21, 25 &  63, 23  \\
          30 &15, 13 & 19, 21 &  45, 31  \\
          40 &14, 14 & 26, 17 &  41, 28  \\
          50 &18, 14 & 31, 28 &  40, 27  \\

  \hline
  \end{tabular}
        \label{tb:dir_err}
\end{table}

\section{Conclusion}
\label{sec:con}
The galaxy distribution traces the background dark matter density. The radio galaxy surveys observe the galaxies relatively at large redshift i.e. $z\sim 1$ over a large sky coverage and therefore the galaxies in these surveys are the potential tracer of large scale matter distribution.  In this work we combine the 
north NVSS and the south SUMSS radio galaxy catalogs and determine the large scale matter distribution. After removing Galactic plane with latitude $\pm 10^\circ$ and other bright locations we get 
$\sim 83\%$ of the sky coverage with number density $\sim 12$ sources/degree$^2$ for flux density cut $S>15$ mJy at 
1.4 GHz. 
The SUMSS is at 843 MHz and NVSS is at 1.4 GHz therefore to combine these surveys we obtain the spectral index 
from overlap region and scale the SUMSS  843 MHz fluxes appropriately to 1.4 GHz. This evidently results as an 
accurate number density match between these two surveys above completeness limits. We further 
calibrate the clustering results from this combined NVSUMSS catalog and find a good match with other works 
\citep{Tiwari:2016adi,Colin:2017}. Except dipole all other multipoles match with $\Lambda$CDM predictions 
and bias values given in \cite{Adi:2015nb} fits very well with observed angular power spectrum. The 
dipole signal remains high and its magnitude and direction matches with previous studies \citep{Tiwari:2016adi,Colin:2017}. 

We study the large scale anomalies i.e. low-$l$ power spectrum, in detail and find that the  quadrupole and octopole signal are nearly consistent with standard $\Lambda$CDM. The quadrupole and octopole directions are on average $46$ and $33$ degree away from dipole. The average angle between quadrupole and octopole is $70$. Although we have large errors in multipole directions due to 
shot-noise. Particularly the octopole directions with different flux cuts are very unstable. Even so the distance between the octopole and quadrupole is large and even with large directional uncertainties we do not find dipole--quadrupole and quadrupole--octopole alignment. However, we can not rule out the dipole--octopole alignment 
with this data. We will have better resolution on these anisotropies and anomalies with upcoming Square Kilometre Array observations \cite{Shamik:2016}. Nevertheless, this work is an independent measure of large scale anomalies observed after CMB. The results in this work 
supports large-scale isotropy which is one of the fundamental assumption in modern cosmology and  thus impose stringent 
constraints on anisotropic cosmological models  and on physical mechanisms introduced to break statistical isotropy on large scales   \citep{Hu:1995,Gordon:2005,Ackerman:2007,Emir:2007,Koivisto:2008,Shamik:2014}. 

We conclude that with available radio galaxy catalogs the large scale multipole directions are random as expected in standard $\Lambda$CDM and with presently available data we do not find CMB like dipole--quadrupole--octopole alignment. The matter distribution at redshift $z\sim 0.8$ is a good match with $\Lambda$CDM (except for dipole power).

\section{Acknowledgments}
We thank Pankaj Jain and Marios Karouzos for a thorough reading of the manuscript and for useful comments on our work.
This work is supported by NSFC Grants 1171001024 and 11673025, and the National Key Basic Research 
and Development Program of China (No. 2018YFA0404503). The work is also supported by 
NAOC youth talent fund 110000JJ01. 
\bibliography{master}
\end{document}